# Functional Intrusive Load Monitor (FILM): A Model-based Platform for Non-Intrusive Load Monitoring System Development

Mengqi Lu, Jinfeng Gao, and Zuyi Li, *Senior Member, IEEE*

*Abstract*—Non-Intrusive Load Monitoring (NILM) is an important application to monitor household appliance activities and provide related information to house owner or/and utility company via a single sensor installed at the electrical entry of the house. It can be used for different purposes in residential and industrial sectors. Thus, an increasing number of new algorithms have been developed in recent years. In these algorithms, researchers either use existing public datasets or collect their own data which causes such problems as insufficiency of electrical parameters, missing of ground-truth data, absence of many appliances, and lack of appliance information. To solve these problems, this paper presents a model-based platform for NILM system development, namely Functional Intrusive Load Monitor (FILM). By using this platform, the state transitions and activities of all the involved appliances can be preset by researchers, and multiple electrical parameters such as harmonics and power factor can be monitored or calculated. This platform will help researchers save the time of collecting experimental data, utilize precise control of individual appliance activities, and develop load signatures of devices. This paper describes the steps, structure, and requirements of building this platform. Case study is presented to help understand this platform.

## I. Introduction

ACCORDING to the U.S. Energy Information Administration (EIA), the energy consumption in the building sector will increase at 0.3%/year in the following years [1]. The Electric Power Research Institute (EPRI) believes that providing customers information of their electricity consumption may reduce the energy usage and help them make monthly budget plan [2]. In this way, the energy insecurity issue in the U.S. can be alleviated [3]. In order to provide electricity consumption and disaggregation feedback to customers, appliance load monitoring (ALM) is one of the solutions.

There are two major ALM approaches which are non-intrusive load monitoring (NILM) and intrusive load monitoring (ILM) [4]. NILM is an important application to monitor household appliances activities and provide related information to house owner or/and utility company via a single sensor installed at the electrical entry of the house. The sensor normally collects voltage and current as the inputs of a NILM system. Through these data, NILM is able to decompose the power usage of individual appliances from overall household power consumption. This procedure is often called energy disaggregation in many publications. The name of "nonintrusive" comes from the meaning of "no need to invade residential area" in the above process. Fig. 1 illustrates the energy disaggregation of the aggregated power in one of our experiments. The other ALM approach is ILM which installs distributed meters on individual devices. This approach can collect various detailed data and relatively accurate information of meter-equipped appliances. In this way, a house owner can execute precise analysis and control of the appliance-based technologies such as power usage pattern, fault detection, and daily activity study in the application of smart homes.

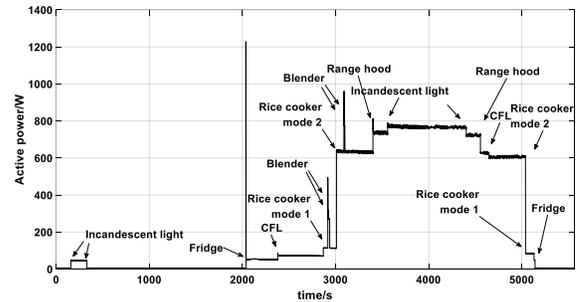

Fig. 1. The aggregated power and power consumption of individual appliances measured at 20 Hz

For most residential buildings, it is impractical to install a sensor at every device in a house due to installation complexity, high cost, privacy issues, technology limitation, etc. Thus, NILM provides a more practical way to achieve multiple purposes via a simple hardware. In his original work [5], Hart believes NILM can be used in energy management, power security, and demand response (DR). In [6], the authors use NILM in commercial buildings for monitoring heating, ventilation and air conditioning (HVAC) system and further indicate its potential in fault detection and power quality monitoring. According to [7], the usage of NILM can be extended in electromechanical units like airplane and vehicles. The authors in [8] develop a new NILM system named real-time electrical appliance recognition (RECAP) which can be used in an industrial setting. Ref. [9] presents a NILM algorithm to detect and identify the charging power consumption of electric vehicles (EV). The authors of [10] point out that NILM is a promising approach to energy management in smart homes.

A common NILM framework usually includes data acquisition, feature extraction, and load identification processes [4]. These processes work step by step to disaggregate meter-level data into appliance-level data. In order to develop an accurate NILM system, the first step is to acquire the aggregated load measurements and ground-truth data at an adequate rate. The selection of sampling rate, electric parameters, recording length, etc. have influences on further detection method selection and result accuracy. The data acquisition process is the foundation of the later steps. However, the inadequate number of desired datasets has limited the development of NILM systems [11]. Existing open datasets mainly include two types, namely the load-level or appliance-level dataset and meter-level dataset. The difference is that to get the former type of data, a sensor is installed at a single device to collect the operation data of that specific appliance while to get the other type of data, a meter is installed at the electric panel to monitor the aggregated data of all the appliances. In the NILM research, the data of individual devices is always used as the training set to extract appliance features or characters and the aggregated data is often employed to develop or test the NILM algorithms in reality. The device-level datasets include ACS-F2 [11], tracebase repository [12], etc. while smart* database [10], BLUED [13], etc. provide aggregated data of home and/or circuits. There are also some datasets that provide the data containing information from both aspects such as REDD [14] and ECO dataset [15].

In most previous NILM research works such as [16], the process of data acquisition has been skipped by using public datasets collected by other researchers. By using these datasets, researchers can save the time of collecting data and focus on evaluating the accuracy of different methods. In the works [17] [18] [6], the development of NILM methods relies on the lab-collected data by the authors themselves. However, some limitations still exist and impede further study. In most cases, the datasets do not provide much information about the tested appliances and some of the events fail to be recorded due to various reasons. In addition, some appliances are not used during the test time. For instance, the air conditioner may not appear in a dataset if the dataset is collected in winter. Besides, there are limited number of electrical parameters being measured in a single dataset which may not be enough for some researches. For example, one of the most frequently used datasets is BLUED [13]. This dataset is designed for event-based NILM approaches and all the events are time stamped and labeled. In the first day of location 1, the phase A of the test house only has 12 appliances in the dataset including 4 unknown appliances and 1 circuit. In this case, researchers need to find their desired data from different types of datasets. On the other hand, the devices with finite operation states such as washing machine and multi-function kettle may produce different events in the dataset. It is difficult for researchers to distinguish each state of the same device from the aggregated data which leads to the loss of some information. Moreover, from the existing datasets, researchers always fail to learn how and why the data of individual appliance is working in that way. This information is important for feature extraction process in the NILM system.

To resolve the above issues, we propose in this paper a model-based platform for NILM system development by building physical circuits of individual appliances in Simscape/MATLAB. This platform aims to be built as a fully controllable model for major electrical appliances. This platform is named Functional Intrusive Load Monitor (FILM) as it is similar to Intrusive Load Monitoring (ILM) in terms of functionality: recording the load profile of every appliance. However, it does that based on model and simulation rather than physical measurement as is the case of ILM. In addition, this platform can generate the data for various electrical parameters of a house and provide a detailed record of the parameters and accompanying events designed by the users of this platform, which is to certain extent like a film that records a movie based on a given screenplay. The FILM platform has the following distinct features:

- It provides an efficient and inexpensive way to monitor the ground-truth of the aggregated electricity consumption data along with appliance-level data which overcomes the issue of data insufficiency in most existing datasets.
- It can provide information on some appliances that are difficult for recording by submeters such as non-plug devices (e.g., lighting and exhaust fan), dedicated circuits devices (e.g., central air conditioning system), and one plug with multiple appliances.
- It can simulate some special cases that are difficult to realize in experiment. Such cases include simultaneous actions of multiple appliances and electric elements breakdown during operation resulting in losing one or more functions in finite state devices.
- It can control all the state transitions of involved appliances. The timings of the state transition actions are labeled precisely which overcomes the issue of missing actions information in prior experiments.
- It can facilitate the development of various algorithms on load signature study, event detection and load identification according to users' requirements. Results of the different algorithms can be evaluated and compared under the same condition provided in the platform.

The remainder of this paper is organized as follows. Section II describes the steps and requirements of building the FILM platform. Section III includes the case study of the individual models and the combined model of the appliances using the FILM platform. Section IV is the conclusion.

## II. CONSTRUCTION OF MODEL-BASED PLATFORM

The proposed FILM platform is a model-based platform, in which the ON/OFF and state transitions of all the involved appliances can be preset by the users of the platform. Meanwhile, multiple electrical parameters such as harmonics and power factor can be monitored or calculated. This platform will help researchers save the time of collecting experimental data, utilize precise control of individual appliance activities, and develop load signatures of devices.

There are four steps to build the FILM platform, namely experimental data collection, data and working principle analysis, individual model construction, and combined model construction, as shown in Fig. 2.

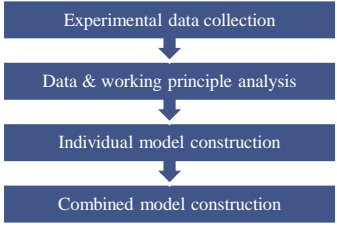

Fig. 2. Flowchart of model-based platform construction

*A. Experimental data collection*

The first step of building the FILM platform is to collect experimental data of individual appliances. In our work, we used Yokogawa WT-3000E precise power analyzer in Fig. 3 to collect the data of typical household appliances at the plug level. About 20 appliances are studied. The measurements are recorded at 20 Hz, including root mean square (RMS) voltage, RMS current, active power, reactive power, power factor, and frequency. According to the user manual [19] of the power analyzer, the data are calculated following the functions listed below.

$$V_{rms} = \sqrt{AVG[u(n)^2]} \quad (1)$$

$$I_{rms} = \sqrt{AVG[i(n)^2]} \quad (2)$$

$$P = AVG[u(n) \cdot i(n)] \quad (3)$$

$$S = V_{rms} \cdot I_{rms} \quad (4)$$

$$Q = s \cdot \sqrt{S^2 - P^2} \quad (5)$$

$$\lambda = P/S \quad (6)$$

where $V_{rms}, I_{rms}$ denote true rms value of voltage and current, respectively; $u(n), i(n)$ denote the $n^{th}$ instantaneous voltage and current, respectively; $S, P, Q$ denote apparent power, active power, and reactive power, respectively; $s$ denotes lead phase (-1) or lag phase (1); $\lambda$ denotes power factor. The frequency is measured using zero crossing detection method on voltage. The above calculation is performed using Average for the Synchronous Source Period (ASSP) on the sampled data at 20Hz.

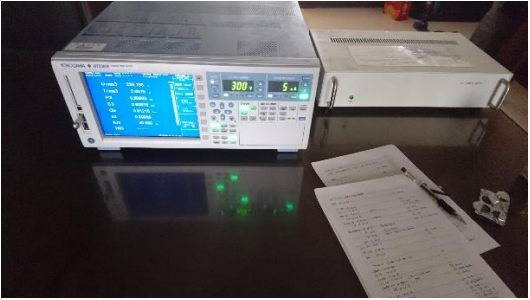

Fig. 3. Yokogawa WT-3000E precise power analyzer

Besides electrical parameters of the individual appliances, the state transition times and corresponding appliance actions are also recorded in the experiment. In order to ensure the accuracy of the data, we have measured multiple sets of data of the same appliance and different brands of the same type of appliance. An example of a split air conditioner activities is shown in Table I.

*B. Data & working principle analysis*

In this second step of building the FILM platform, statistics algorithm is applied to the data of individual appliances. The working principle and possible states of the appliances are analyzed. The above works aim to provide a clear understanding of the target appliance so that the circuit model can maintain the main function and special features of the original appliance during simulation. The analysis includes

a. Calculating the mean value and standard deviation of electrical parameters in each state:

$$Mean = \frac{1}{n}\left(\sum_{i=1}^{n} x_i\right) \quad (7)$$

$$Std = \sqrt{\frac{1}{n-1}\sum_{i=1}^{n}|x_i - Mean|^2} \quad (8)$$

where $n$ is the number of selected samples in steady state of the appliance, and $x_i$ is the $i^{th}$ selected electrical parameters (P, Q, I, and λ) in steady state.

b. Discovering geometric features of the data such as the appearance of spikes, possible states, cycles, and duration.

c. Determining the working principle and major components of the appliances.

Table II shows a statistics analysis example of the air conditioner in cooling and fan mode.

TABLE I
RECORD OF A SPLIT AIR CONDITIONER ACTIVITIES IN THE EXPERIMENT

| State | Absolute Time (hh:mm:ss) | Relative Time (s) | User Action |
|---|---|---|---|
| Start | 11:25:00 | 0 | |
| ON/Cool | 11:25:18 | 18 | Speed=5, T=21 |
| - | 11:51:40 | 1500.6 | Speed=3 |
| - | 11:56:00 | 1860 | T=16 |
| … | | | |
| - | 14:05:48 | - | T=21 |
| - | 15:04:00 | 13139.6 | Speed=5 |
| OFF | 15:06:00 | - | |
| Total time | 221.0008 min | 13260.05 | |

The first column shows the working state of the air conditioner. The second and third columns show the absolute time in the experiment and the corresponding relative time to the start of the experiment when the state changes or user action occurs. The last column shows the user action which includes fan speed control (1 to 5, 5 for highest) and temperature (T) selection in centigrade (C).

TABLE II
STATISTICS ANALYSIS OF A SPLIT AIR CONDITIONER

COOLING MODE (WITH FAN SPEED=1, 3, 5):

| | I (A) | P (W) | Q (Var) | PF (%) |
|---|---|---|---|---|
| Speed=1 | 4.732±0.074 | 1097.25±17.67 | 210.88±9.95 | 98.20±0.19 |
| Speed=3 | 4.914±0.096 | 1135.46±23.13 | 218.57±11.67 | 98.19±0.21 |
| Speed=5 | 4.897±0.078 | 1133.73±19.09 | 222.31±11.21 | 98.13±0.21 |

FAN MODE (WITH FAN SPEED=1, 2, 3, 4, 5):

| | I (mA) | P (W) | Q (Var) | PF (%) |
|---|---|---|---|---|
| Speed=1 | 142.99±1.634 | 14.32±0.31 | 30.99±0.29 | 41.94±0.51 |
| Speed=2 | 150.06±1.484 | 15.88±0.30 | 32.03±0.26 | 44.41±0.45 |
| Speed=3 | 156.15±1.600 | 17.42±0.34 | 32.96±0.28 | 46.71±0.49 |
| Speed=4 | 160.38±1.266 | 18.88±0.28 | 33.33±0.21 | 49.29±0.41 |
| Speed=5 | 163.75±1.017 | 20.45±0.23 | 33.27±0.17 | 52.37±0.35 |

*C. Individual model construction*

The purpose of this third step of building the FILM platform is to construct a circuit model for each experimented electrical appliance. Based on the information in the previous step, the circuit model is constructed to meet two requirements. The first requirement is that the model must contain the major components and can achieve the main function of the device under study. This requirement aims to make the circuit model a general one which can represent a wide range of appliances of the same type. For refrigerators, regardless of the brands, its major power consumption element is often compressor and its main function is to keep the inner temperature at the pre-set value. Then the model should mainly contain a compressor circuit and use temperature as control signal.

The other requirement is that the model should work following the working principles of the appliance. This principle is the way how the appliance works. It can be discovered from the waveforms of each electrical parameter during the working condition. Fig. 4 shows the waveforms of a tested blender.

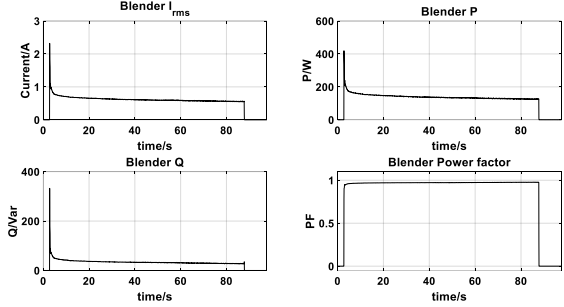

Fig. 4. RMS current, active power, reactive power, and power factor of the tested blender during working condition

One of the important working principles is the operation states. Depending on the number of possible working states, the household appliances can be classified into different categories which are shown below [1-2]. It should be noted that in the model construction process, the meaning and the classification rule of the first two types of appliances is slightly different from that in previous work.

a. ON/OFF appliance. This type of appliance only has two possible states: either ON or OFF. Although some multifunction appliances have multiple switches or buttons on the panel, their control method is that they only adjust the working time of the appliances while the working condition does not change. In this way, the appliance like toaster and multifunction kettle still goes to this category. This is slightly different from preview work.

b. Finite state machine (FSM). An appliance belonging to this type usually has distinct multiple operation states and each state is repeatable under the control of users. The power consumption and current waveforms among each state usually show significant difference.

c. Continuously variable devices (CVD). This type of appliance does not have clear states or can be seen as having infinite states. Its power consumption changes with time or user action. Moreover, the power and current waveforms usually change beyond normal oscillations observed in other appliances and its startup and shutdown powers are often different.

d. Permanent consumer devices. The always-on devices such as modem, router, and telephone belong to this category. In some cases, some parts of other type of devices are also treated in this category, for instance, the LED digital clock on the range hood. These devices usually have low power consumption and working current compared to other types of appliance and are often treated as background power in some NILM algorithms.

*D. Combined model construction*

The last step of building the FILM platform is to construct a combined model for all experimented electrical appliances. Most electrical appliances in residential houses in the U.S. are wired in parallel which means the appliances work independently from each other. The overall power consumption of the house equals the sum of power consumption of all the connected appliances which can be express as follows.

$$P_o = \sum_{i=1}^{n} P_i + E \qquad (9)$$

where $P_o$ is the active power consumption of the house; $P_i$ is the $i^{th}$ electrical appliance; $n$ is the number of electrical appliances; $E$ is the total loss.

In the FILM platform, multiple individual appliance models are connected in parallel as a combined model to simulate the real house. An example is given in the case study section.

### III. CASE STUDY

As described in the previous section, the model is built based on the real household electric appliance and is modified to cover more devices of the same type. After the individual models have been constructed, users can assemble their desired appliance models together to build the combined model of a house. In this way, the aggregated data and ground-truth data can be measured from simulation. The example of individual and combined models will be presented in this section. Although the standard wire voltage and frequency is 220V and 50Hz in the testing lab, the power analyzer measurement of the wire voltage was around 235V. Therefore, the source in the model is set to 235V, just to be consistent with the measurement.

*A. Individual appliance model*

The tested appliance is a Haier 215L BCD-215KS energy-saving fixed-frequency refrigerator including a 78L freezer. The compressor will operate ON and OFF to keep the inner temperature matching the preset value. When the compressor starts from shutdown status, high inrush current is needed to start the motor. The changes of electrical parameters in steady state of each cycle result from the PTC thermistor resistance changes in the motor circuit.

Fig. 5 shows the circuit model built for the refrigerator. There are two sections in this refrigerator circuit model namely door light circuit and compressor motor circuit. $L_1$, $R_1$ and switch $S_1$ form a 1.5W LED light. The remaining elements include a single-phase resistance start capacitor run motor (RSCR) which is the most commonly used motor type in refrigerator compressor. $L_{st}$ and $R_{st}$ represents the start winding, while $L_r$ and $R_r$ stands for the run winding. $C_{run}$ is the running capacitor and PTC is positive temperature coefficient (PTC) thermistor. When the motor first starts, the low temperature PTC thermistor has low resistance that shorts $C_{run}$ and most current will bypass the PTC element. This circuit provides high starting current to the motor. After that, the current continues flowing through thermistor raising its temperature and its resistance decreases first and then increases exponentially to thousands of ohms. In this way, the PTC thermistor can be seen as an open circuit, and $C_{run}$ is connected to the circuit gradually until the circuit reaches the rated working condition. When the target temperature is reached, the compressor will stop working and the whole circuit barely consumes any power. The thermal signal in the model simulates the temperature changes over PTC thermistor. Users can define the working time of the compressor and door light by using control signal subsystem. The input monitor subsystem is able to monitor the input electrical parameters such as voltage, current, active/reactive power etc. which can be preset by users.

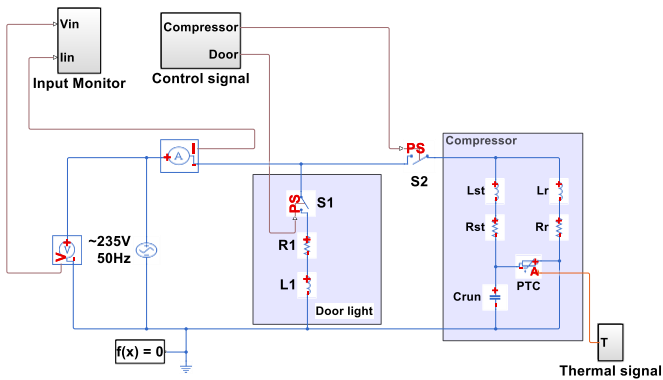

Fig. 5. Model of fixed-frequency refrigerator

The designed test for this model is 17000s and the test starts when the compressor is shutdown. The entire test is described in Table III.

TABLE III
DESIGNED SIMULATION TEST OF REFRIGERATOR MODEL

| State | Relative Time (s) | User Action |
|---|---|---|
| Start/Compressor OFF | 0 | |
| Compressor ON | 24.25 | |
| | 144.15 | Door open |
| | 195.5 | Door closed |
| Compressor OFF | 10,334 | |
| | 10,583.75 | Door open |
| | 10,612.25 | Door closed |
| | 12,055.4 | Door open |
| | 12,066.45 | Door closed |
| Compressor ON | 12,828.6 | |
| Compressor OFF | 16,644.8 | |
| End | 17,000 | |
| Total time | 17,000 | |

Figs. 6 and 7 show the measurements data and simulation results of the model. The parameters include RMS current, active power, reactive power, and power factor.

In the plots of RMS current and active power, the spikes at the turn-on period is due to the compressor motor behavior. The long transient afterwards results from the characteristics of the PTC elements. The cycles of the turn-on and turn-off show the working patterns of the refrigerator. The small rise and fall during the working period are caused by the door light activities. The above features are also shown in the measurements of tested appliance. It should be noted that the working time of the compressor is not constant since it is decided by the inner temperature and coefficient of performance (COP) of the refrigerator.

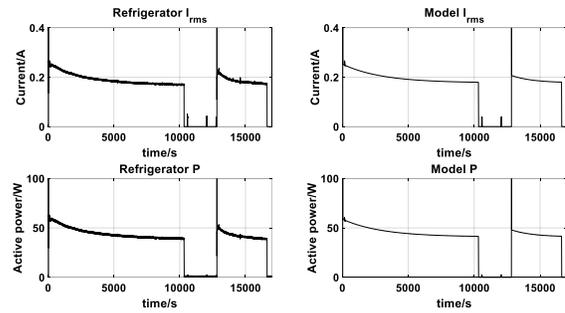

Fig. 6. RMS current and active power plots of measurements and model simulation

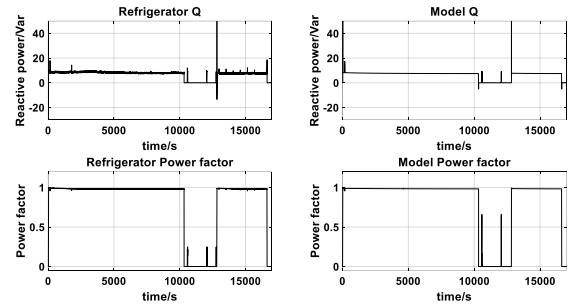

Fig. 7. Reactive power and power factor plots of measurements and model simulation

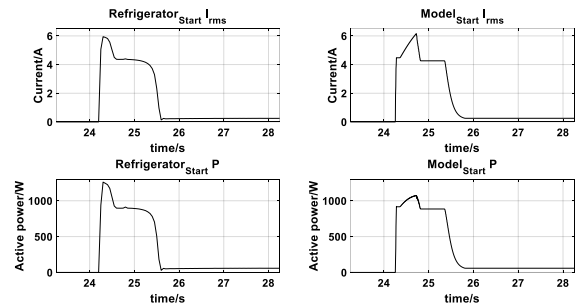

Fig. 8. Transient plots of RMS current and active power of measurements and model simulation

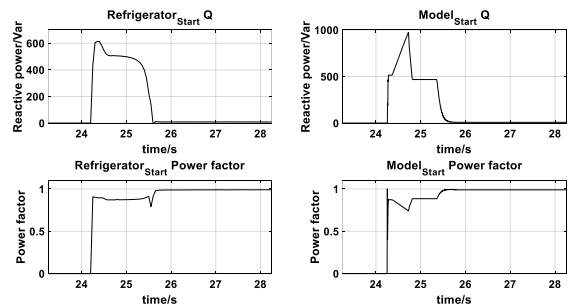

Fig. 9. Transient plots of reactive power and power factor of measurements and model simulation

Figs. 8 and 9 show the transient plots of measurement data and simulation results of the model when the compressor turns on. The parameters include RMS current, active power, reactive power, and power factor.

The measurements of the tested appliance show a significant spike when the compressor turns on, and then return to flat high values before they decline to rated value. In the turn-on transient plot of the model, it successfully reproduces the trends and spikes of the electrical parameters. The transient characters are important features for further event detection and load identification in some NILM algorithms.

The accuracy of the model is evaluated by using both percentage error and correlation coefficient. The former metrics is the difference between model result and actual measurement in percentage which is defined in (10). This error aims to evaluate the results differences. However, (10) cannot reflect the real situation when the measurement value is small or zero, for which special consideration is made to make the evaluation more reasonable. For example, when the actual measurements are within certain small values (0.001A for RMS current, 2W and 2Var for active and reactive power, 0.01% for power factor), and the differences between the actual measurement and model results are also falling in this range, the percentage error is set to 0 instead of infinity or a large value.

The second metrics is to describe the closeness of the waveforms between model result and actual measurement. The correlation coefficient showing in (11) [20] is able to show movements and trends of samples.

$$E = \frac{|x_i - y_i|}{|x_i|} 100\% \tag{10}$$

$$r = \frac{n \sum x_i y_i - \sum x_i \sum y_i}{\sqrt{n \sum x_i^2 - (\sum x_i)^2} \sqrt{n \sum y_i^2 - (\sum y_i)^2}} \tag{11}$$

where $E$ denotes the percentage error; $i$, $x_i$ and $y_i$ denote the $i^{th}$ value in actual measurement and model result, respectively; r is the correlation coefficient; $n$ is the number of measurements. Fig. 10 shows the plots of modified percentage error of the refrigerator model. Table IV shows the results of correlation coefficients for different electrical parameters of measurements and model results.

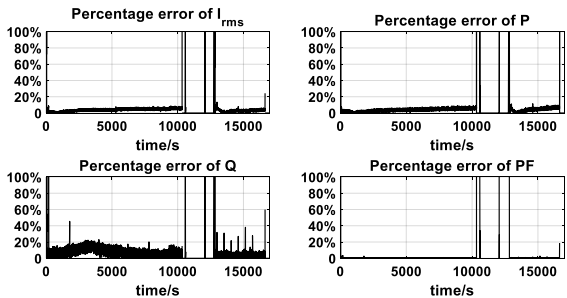

Fig. 10. Percentage error of RMS current, active power, reactive power and power factor of refrigerator between measurements and model simulation

TABLE IV
CORRELATION COEFFICIENT OF MODEL RESULTS AND MEASUREMENTS

| Parameters | $I_{rms}$ | P | Q | PF |
|---|---|---|---|---|
| Correlation r (%) | 96.57 | 97.43 | 86.56 | 99.98 |

From Fig. 10, the percentage errors of the above electrical parameters remain at a low level. The relatively large errors mainly appear at the turn-on point of the compressor due to two reasons. The first is the value difference of the inrush current. The model does not have the exact turn-on shape as that in the measurement since lags may exist. However, according to the results in Table IV, the model can reproduce the trend and most of the turn-on transient in the model shown in Figs. 8 and 9. Because the inrush value is about 6A in the experiment, even a minimum difference may lead to a high percentage error. The second reason is the voltage fluctuation and background noise in the experiment. Since the test refrigerator has high power factor, its reactive power is small and easy to be influenced by the background power produced by noise and voltage fluctuation. Most errors of reactive power and power factor are caused by this factor. The accuracy can be further improved by considering voltage noises and fluctuations in the model and the platform as discussed in Section D.

*B. Combined household model*

In this section, a combined model is built to simulate a residential house which contains different types of appliances. This model contains 7 appliances including a coffee machine, a toaster, a water kettle, a range hood, an incandescent light, a refrigerator, and a dimmable light. The coffee machine and kettle can turn off automatically after the drink is ready. The toaster has 7 different time settings. The above appliances can simulate type I appliance with automatic control function. The range hood has a 2-speed fan and a permanent-on digital clock which can model type II and type IV appliance. The light intensity of the dimmable lamp is controlled by a dimmer switch which belongs to type III appliance. Fig. 11 shows the combined model.

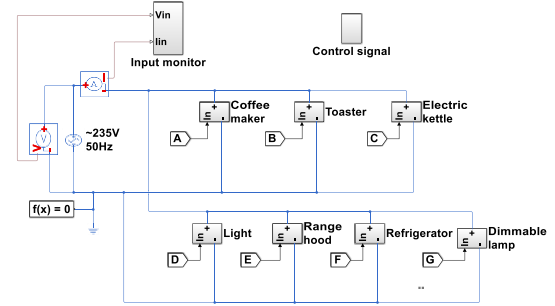

Fig. 11. Combined model of different types of appliances

The subsystems are the individual models of corresponding appliances. The signal acceptance blocks ('from' blocks) such as A and B receive the control signal from the control signal subsystem. The self-switch-off function of the coffee maker and kettle is achieved by automatic switch block by using thermal equation shown below.

$$t = \frac{Q}{P} = \frac{cm(T - T_0)}{P \cdot \eta} \tag{12}$$

where $t$ is the switch off time; $c$ is the specific heat capacity which is $4.2 \times 10^3 \text{J}/(\text{kg} \cdot °\text{C})$ for water; $m$ is the mass of water which can be set in the control block; $T$ is the boiling point which is $100°\text{C}$ at standard atmosphere pressure; $T_0$ is the initial water temperature which is preset to room temperature at $25°\text{C}$; $P$ is the rated power of the appliance; $\eta$ is the heat transfer rate.

The designed simulation time for this model is about 29.5 minutes which has 16 appliance activities. In this simulation, all the appliance activities and corresponding times are the same as those in the measurements. The detailed appliance activities are shown in Table V.

TABLE V
RECORD OF APPLIANCES ACTIVITIES IN THE EXPERIMENT

| State | Absolute time (hh:mm:ss) | Relative Time (s) | User Action |
|---|---|---|---|
| Start | 21:45:30 | 0 | |
| Incandescent ON | 21:46:43 | 73.5 | |
| Refrigerator ON | 21:49:47 | 257.1 | |
| Coffee machine ON | 21:50:53 | 323.35 | |
| Toaster ON | 21:53:27 | 476.7 | |
| Coffee machine OFF | 21:55:54 | 623.7 | |
| Toaster OFF | 21:56:10 | 639.75 | |
| Range hood ON | 21:56:45 | 674.35 | Speed=3 |
| Range hood | 21:57:38 | 728.4 | Speed=2 |
| Kettle ON | 21:59:00 | 810.15 | |
| Range hood OFF | 22:00:07 | 876.85 | |
| Kettle OFF | 22:03:15 | 1064.9 | |
| Incandescent OFF | 22:04:49 | 1158.85 | |
| Refrigerator OFF | 22:09:27 | 1436.55 | |
| Dimmable light ON | 22:10:20 | 1490 | Dimmest |
| | 22:11:12 | 1542 | Brightest |
| Dimmable light OFF | 22:13:50 | 1701.05 | |
| End | 22:15:00 | | |
| Total time | 29.5 min | 1770 | |

Figs. 12 and 13 show the RMS current, active power, reactive power and power factor of the measurements and the simulation model.

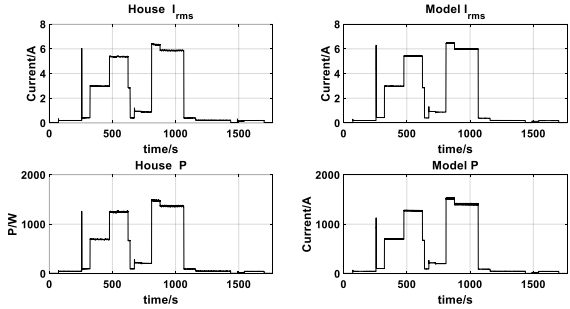

Fig. 12. RMS current and active power plots of measurements and model simulation

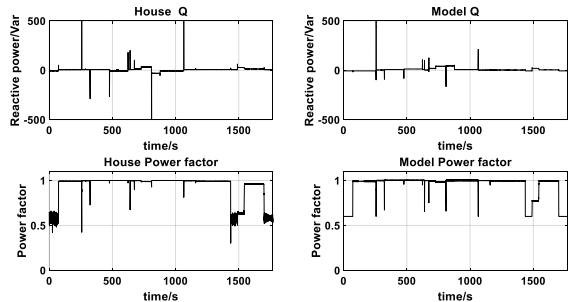

Fig. 13. Reactive power and power factor plots of measurements and model simulation

From comparison of the above figures, the simulation results are able to show the power and current changes and most of transient spikes in the tested period.

## C. Different brands of the same type of appliances

As discussed in section II.C, one of the advantages of the FILM platform is that a model should be able to represent the same type of appliance from different brands. By using the same model in Fig. 5 and adjusting the parameters of the model, the simulation results of the other device are shown in Figs. 14 and 15. The tested appliance is a Meiling BCD-109ZM2 109L refrigerator including a 37L freezer. During the experiment, the cooling level is set at 5 out of 7.

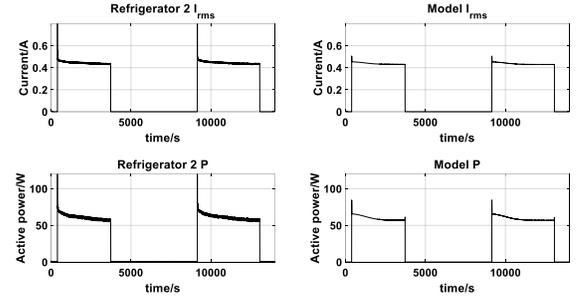

Fig. 14. RMS current and active power plots of measurements and model simulation

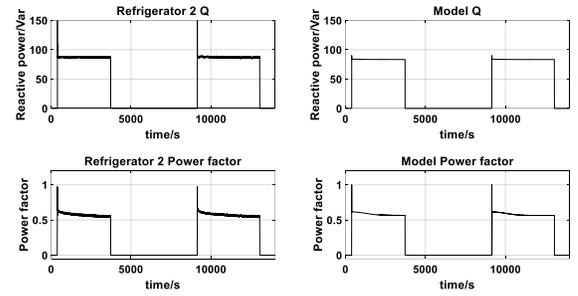

Fig. 15. Reactive power and power factor plots of measurements and model simulation

From the comparison of the results between the model and measurements, the model can reproduce the working cycle of the compressor and changing trends of the above parameters in each cycle. The spikes in each parameter when the compressor turns on are also shown in the simulation results. One limitation is that the magnitudes of the current and reactive power spikes are not as significant as that in the measurements. But these spikes still can be observed.

Although the measurements of the two refrigerator brands shown in Figs. 6 and 7, Figs. 14 and 15 have significant differences in working time, RMS current value in turn-on transient and steady state, power consumption, and power factor, they can still be simulated by using the same generalized model in Fig. 5 with only an adjustment of parameters. This is one of the advantages of the proposed FILM platform.

## D. Model with noise

By using the FILM platform, special cases like voltage noise and drop which are difficult to be monitored or recorded for traditional methods can be simulated. Figs. 16 and 17 show the results of a kitchen model at breakfast time. The tested appliances include a coffee machine, a toaster, and a water kettle. The voltage noise and drop are shown in these figures.

The simulation results depict the voltage oscillation caused by noise and small voltage drop during the working period of the appliances.

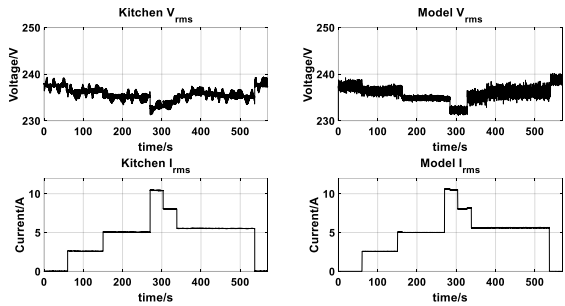

Fig. 16. RMS voltage and RMS current plots of measurements and model simulation

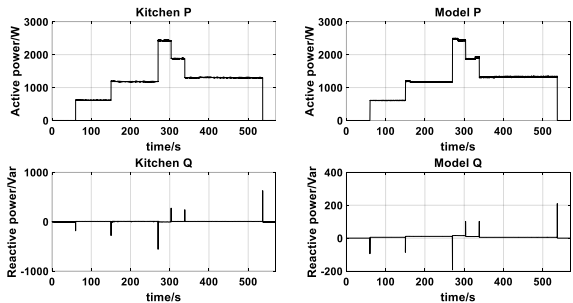

Fig. 17. Active power and reactive power plots of measurements and model simulation

## IV. CONCLUSION

In this paper, we build a new platform for NILM system development namely FILM. Rather than using the existing public datasets, this model provides a controllable way for researchers to join in the data collection process. Both aggregated data and ground-truth data can be monitored under the proposed FILM platform. Extra electrical parameters can be measured and calculated in the platform. Moreover, additional conditions such as signal noise, voltage drop, and device broke down can be added in the platform for other usage. Load signature study, event detection and load identification algorithms can be developed by using the data produced by this platform. Moreover, FILM provides a platform for performance evaluation of different NILM systems under the same condition.

There are still some limitations of the current FILM platform. The uploaded rate of the original measurements is 20Hz. The ability of FILM to simulate the data in higher rate remains to be further developed. From the evaluation results in III. C, the accuracy of the models cannot reach 100%, which means the simulated data is not a complete refection of the actual data. However, caution should be used when trying to balance the accuracy and the generality of the model. Furthermore, when developing NILM related algorithms based on this platform, the robustness and load signature selection should be carefully considered.